# Low Temperature Structural Phase Transition in the Perovskite Ba$_2$CaMoO$_6$

*Loi T. Nguyen,[1,2] Robert J. Cava,[2] Allyson M. Fry-Petit[1]\**

[1]Department of Chemistry and Biochemistry, California State University, Fullerton, California 92831, United States

[2] Department of Chemistry, Princeton University, Princeton, New Jersey 08544, United States

*Corresponding author

**Abstract**

Ba$_2$CaMoO$_6$ was synthesized by solid state method. The crystal structure adopts cubic *Fm*-3*m* space group at room temperature with lattice parameters of 8.378231(5) Å. Upon cooling, Ba$_2$CaMoO$_6$ was determined to have a structural phase transition to tetragonal *I*4/*m* (a=5.905763(6) Å and c=8.38817(1) Å) around 200 K. The phase transition was probed structurally by synchrotron and neutron diffraction and thermodynamically by specific heat and differential scanning calorimetry measurement. This structural phase transition will deepens our understanding of the perovskite family especially the formation of perovskites that break corner sharing networks.

**Introduction**

Perovskites became intensively studied after the discovery of dielectric and ferroelectric properties in the perovskite BaTiO$_3$.[1–9] The crystal structure often governs the material properties of a solid, hence if the structure can be altered under temperature or pressure, the material can be tuned to have the structure and thus properties needed for the application of interest that is dependent on an external stimulus. For example, $Sr_3WO_6$ undergoes several temperature dependent phase transitions as studied by electron diffraction, synchrotron x-ray diffraction and neutron powder diffraction. The monoclinic γ-$Sr_3WO_6$ structure (space group $Cc$) was observed at the temperature above 470 K, and gradually converted to a triclinic β-$Sr_3WO_6$ phase (space group $C1$) at room temperature.[10] $Sr_3WO_6$ is of great interest in this study as its structure displays large degree octahedral tilting that breaks the prototypical corner sharing network in both the γ and β phases and is thus considered a noncooperative octahedral tilted (NCOT) perovskite.

In perovskites the Goldschmidt tolerance factor (τ)

$$\tau = \frac{r_A + r_X}{\sqrt{2}(r_B + r_X)}$$

where $r_A$, $r_B$, and $r_X$ are the ionic radii of $A$, $B$, and $X$ respectively, can be used to predict if the structure will undergo structural distortions from the cubic phase to accommodate the differing sizes in the cation. Ideal cubic perovskite structures usually have τ in range 0.9-1. In the case of a relatively large $A$-site cation and relatively small $B$-site cation, the tolerance factor becomes larger than 1 and the $BO_6$ octahedra are face-sharing or edge-sharing instead of corner-sharing and a hexagonal perovskite is formed. Conversely, when the tolerance factor is in the range 0.7-0.9, distortion of $BO_6$ octahedra leads to cooperative octahedral tilting (COT) that accommodates for a small $A$-site cation. Even smaller tolerance factors results in the stabilization of the $LiNbO_3$ structure type. In $Ba_2CaMoO_6$, the tolerance factor is calculated to be 0.98 implying that it maybe cubic at all temperatures. However, in other perovskites, such as the aforementioned NCOT perovskite, $Sr_3WO_6$, as well as $K_3WO_3F_3$, $Rb_3WO_3F_3$, $K_3AlF_6$, $Rb_2KCrF_6$, $Rb_2KGaF_6$, and $CsInCl_3$ all possess relatively high tolerance factors of 0.94, 0.96, 0.95, 0.95, 0.93, 0.92, and 1.0 respectively.[10–14] In all of these compounds the NCOT tilting structure cannot be adequately predicted by tolerance factor alone. One of the additional parameters that has been proposed as a predictor of NCOT compound formation is the difference of ionic radii of the *A*, *A'*, and *B* cations (termed ΔIR counter ions), which is relatively small for all of the pervious compounds listed with values of either 0 or 0.34. Understanding the role of ΔIR of counter ions is of interest as it has been proposed that NCOT structures are stabilized due to *A*-site ions becoming more *B*-site like in their coordination and vice versa.[11]

Many NCOT compounds undergo complex structural transitions as the result of rotation around different axis as a function of temperature, like COT all are proposed or observed to adopt the cubic perovskite structure at high temperatures. The temperature dependent structure of $Ba_2CaMoO_6$ was of interest in probing the role of ΔIR counter ions in the formation of NCOT structures. There are several geometric requirements that have been proposed as important in NCOT perovskites: rock salt ordering, tolerance factor, difference in the ionic radius of the *B* and *B'*-site, and ΔIR of counter ions.[10,11,15] The difference in the oxidation state $Ca^{2+}$ and $Mo^{6+}$ is four which promotes NCOT rock salt ordering, which has been seen in all known NCOT materials.[a][16] The tolerance factor is relatively large at 0.98, but in the range of values observed for NCOT compounds, and significantly above what is observed for compounds that adopt the $LiNbO_3$ structure. $Ba_2CaMoO_6$ has a difference in the ionic radius of *B* and *B'* cations of 0.41, which is expected to be in the range of 0.50 and 0.66 Å for NCOT oxides, and therefore is lower then expected for NCOT compounds.[15] Furthermore, relatively large ΔIR of counter ions, 0.61 Å, which is above the highest value of 0.34 Å for the list of NCOT compounds given above. This work probed if the NCOT structure is observed in a system where the difference in the ionic radius of *B* and *B'* was low while the difference in ΔIR of counter ions was high. To that end both synchrotron and neutron diffraction data at were collected for temperatures ranging from 300 K to 10 K and found that the crystal structures adopt the ideal cubic, *Fm*-3*m* (225), at high temperature and cooperative tilting, *I*4/*m* (87), at all temperatures probed below 300 K. Furthermore, the transition temperature from cubic to cooperative tilting was found to be around 200 K by heat capacity and differential scanning calorimetry.

---

[a] Note that in $CsInCl_3$ the rock salt ordering is due to charge ordering of the indium *B*-sites.

**Materials and Methods**

*Synthesis*

$Ba_2CaMoO_6$ was successfully synthesized by solid state method. Stoichiometric amounts of high purity $BaCO_3$ (Alfa Aesar, 99.9%), $CaCO_3$ (Alfa Aesar, 99.5%), and $MoO_3$ (Stream Chemicals, 99.5%) were ground in an agate mortar and pestle for approximately 20 minutes. The sample was heated in an open alumina crucible for 24 hours at 1073 K followed by several additional heatings at 1373 K for 12 hours, with intermediate manual grinding between each heating. The purity of the sample was probed after each heating using a Rigaku Miniflex powder x-ray diffractometer equipped with a copper source ($\lambda=1.54$ Å) and a variable slit aperture. Synthetic attempts with oxides in place of the carbonates resulted in secondary phases.

*Structural Analysis*

Low temperature powder x-ray diffraction was collected on a Bruker D8 diffractometer (40 kV, 50 mA, sealed Cu x-ray tube) with a silicon drift detector and equipped with a closed cycle helium Oxford Cryosystems PheniX chiller. Synchrotron data was collected at 100 K and 300 K on beamline 11-BM at the Advanced Photon source at Argonne National Laboratory at a wavelength of 0.412728 Å. Time-of-flight neutron data was collected on POWGEN powder diffractometer. Banks 2 and 4 were used for refinement. All Rietveld refinements were performed in TOPAS Academic software package.[17]

*Thermodynamic Analysis*

Heat capacity measurements were performed on 6.1 mg of $Ba_2CaMoO_6$ powder pressed into a pellet and annealed at 1373 K overnight. The pellet was mounted in a platform of heat capacity puck by using N-grease. The heat capacity was measured from 300 to 100 K by a Quantum Design Physical Property Measurement System (PPMS) DynaCool equipped with a heat-capacity option. Differential scanning calorimetry (DSC) was collected on a 1.1 mg sample of $Ba_2CaMoO_6$ in a Concavus Pan Al with a pierced lid. The DSC cooling curve was recorded on a NETZSCH DSC 214 with at the rate 2.0 K/min from 298 K to 88 K.

**Results and discussions**

To probe for the presence of a phase transition below room temperature powder XRD was collected at room temperature, 100 K and 12 K. As can be seen in the insets of Figure 1 a splitting of several peaks are observed in the 100 K and 12 K data relative to the room temperature data. This indicates a decrease in the symmetry of the structure upon cooling. Initial Rietveld refinements of the room temperature powder XRD were well fit by a cubic *Fm-3m* model, however due to very fine splitting that can be present in materials that adopt NCOT both neutron and synchrotron data were collected on $Ba_2CaMoO_6$.

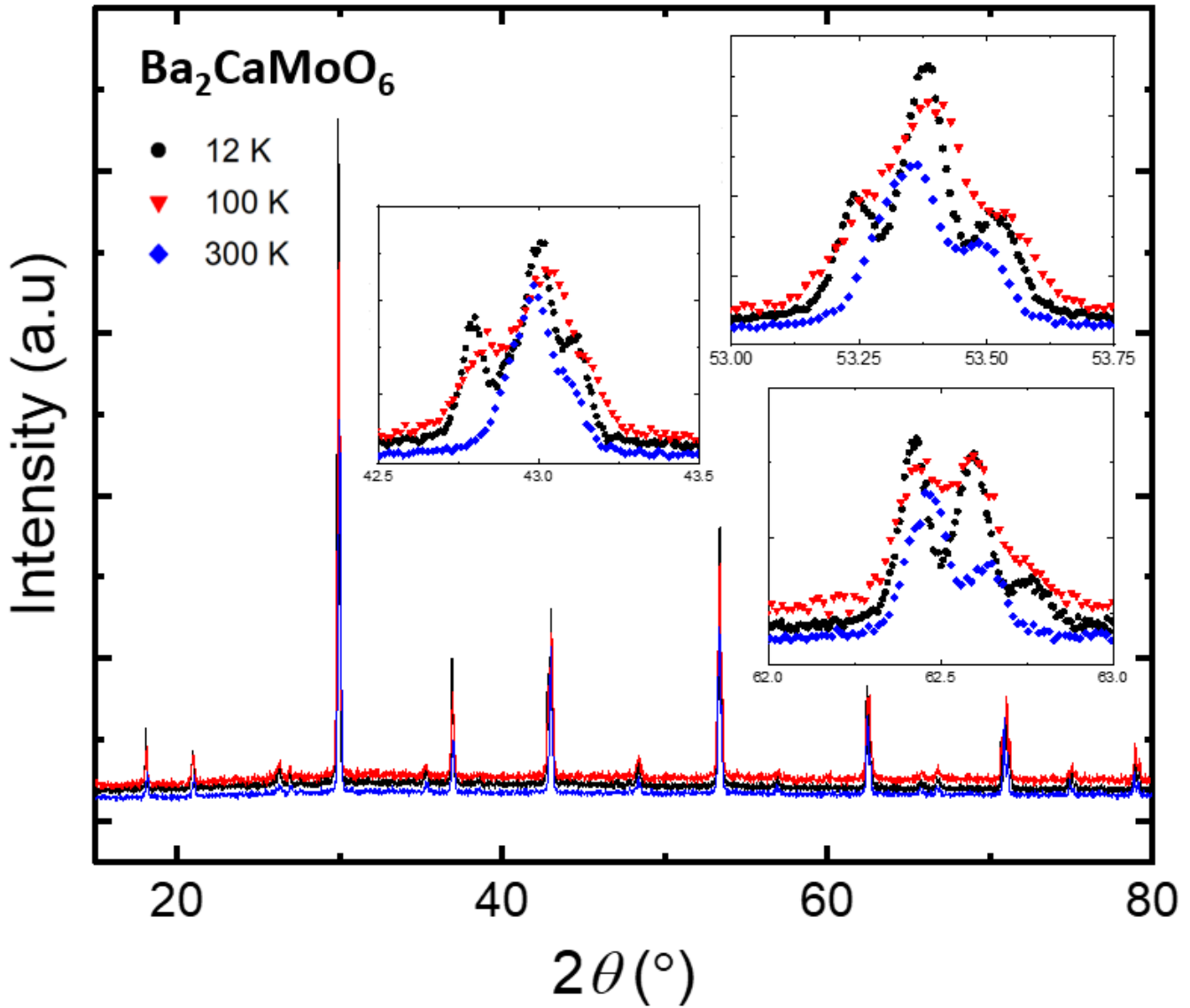

Figure 1. Powder x-ray diffraction of $Ba_2CaMoO_6$ at 12 K (black circles), 100 K (red triangles) and 300 K (blue diamonds). Peak splitting indicates that the crystal structure adopts lower symmetry space group at low temperature as discussed above.

To further probe the temperature at which the structural phase transition occurs the heat capacity was measured. As show in **Error! Reference source not found.**, a weak anomaly was observed in $C_p(T)$ near 200 K, indicating the small loss of entropy due to the structural phase transition in $Ba_2CaMoO_6$. The inset of **Error! Reference source not found.** shows the first derivative of $C_p(T)$ and more clearly shows that the phase transition is occurring around 200 K. This subtle structural phase transition at high temperature is similar to the case of $Ba_5AlIr_2O_{11}$, which has a structural phase transition temperature near $T_S = 210$ K.[18]

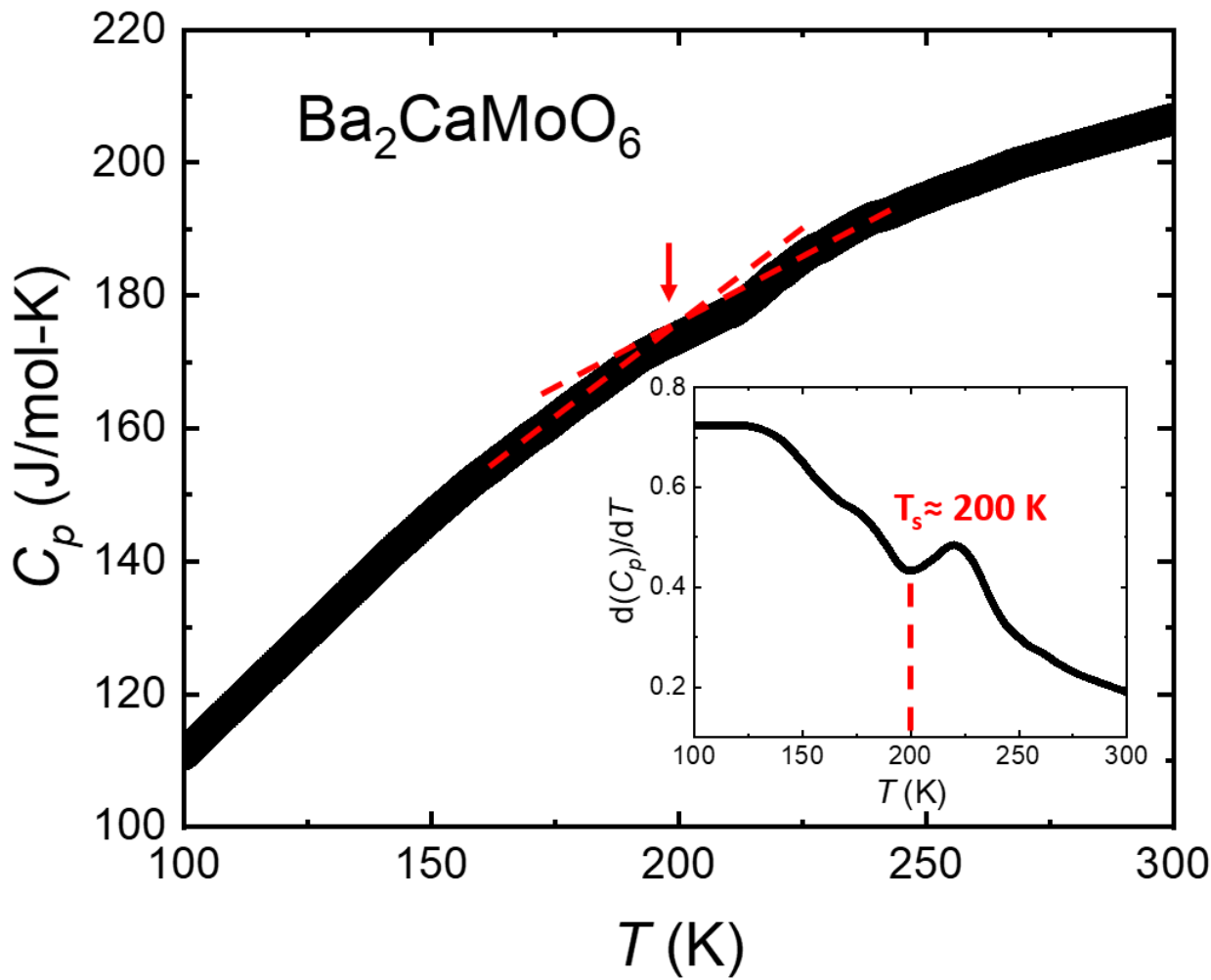

Figure 2. Heat capacity of $Ba_2CaMoO_6$ pellet sample measured from 300 to 100 K under zero field. The change in slope of heat capacity indicates phase transition at around 200 K, confirmed by the first derivative of $C_p$ in the inset.

DSC can be used to study structural changes as well due to the fact upon heating or cooling most phase transitions contain a substantial amount of enthalpy, however in structural transitions such as tilting of octahedra these transitions can be relatively small and necessitate slow scan rates. The structural phase transition at around 200 K in Ba$_2$CaMoO$_6$ was captured by the DSC measurement shown in **Error! Reference source not found.**. On cooling from 298 K to 88 K, the characteristic peak at 200 K indicates the phase transition from cubic to tetragonal symmetry as discussed above. This temperature agrees well with slope changes at around 200 K in the heat capacity measurement.

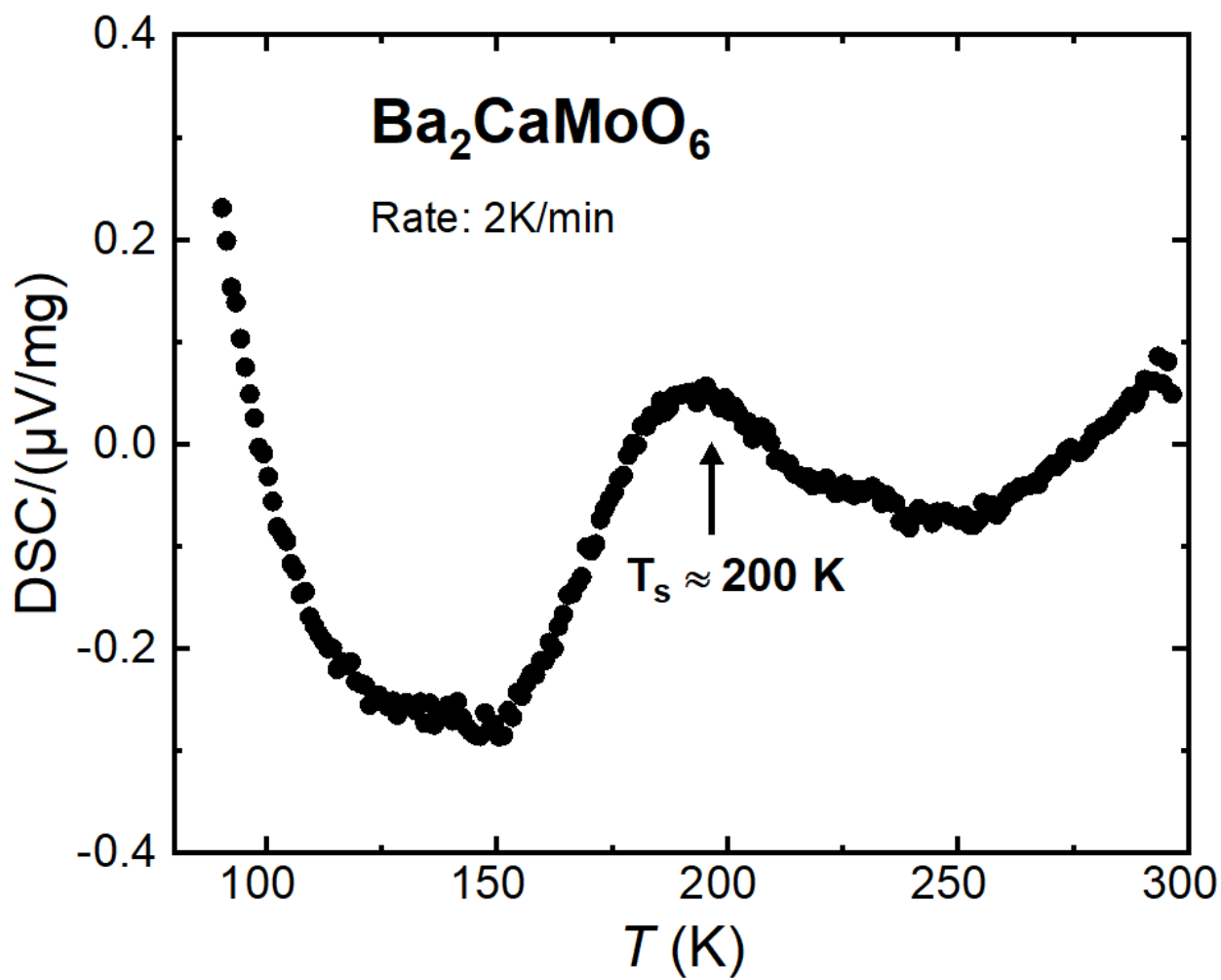

Figure 3. DSC curve of Ba$_2$CaMoO$_6$ was recorded from 298 K to 88 K at the cooling rate of 2.0 K/min. The arrow marks the temperature at around 200 K where the structural phase transition was observed in heat capacity measurement as well.

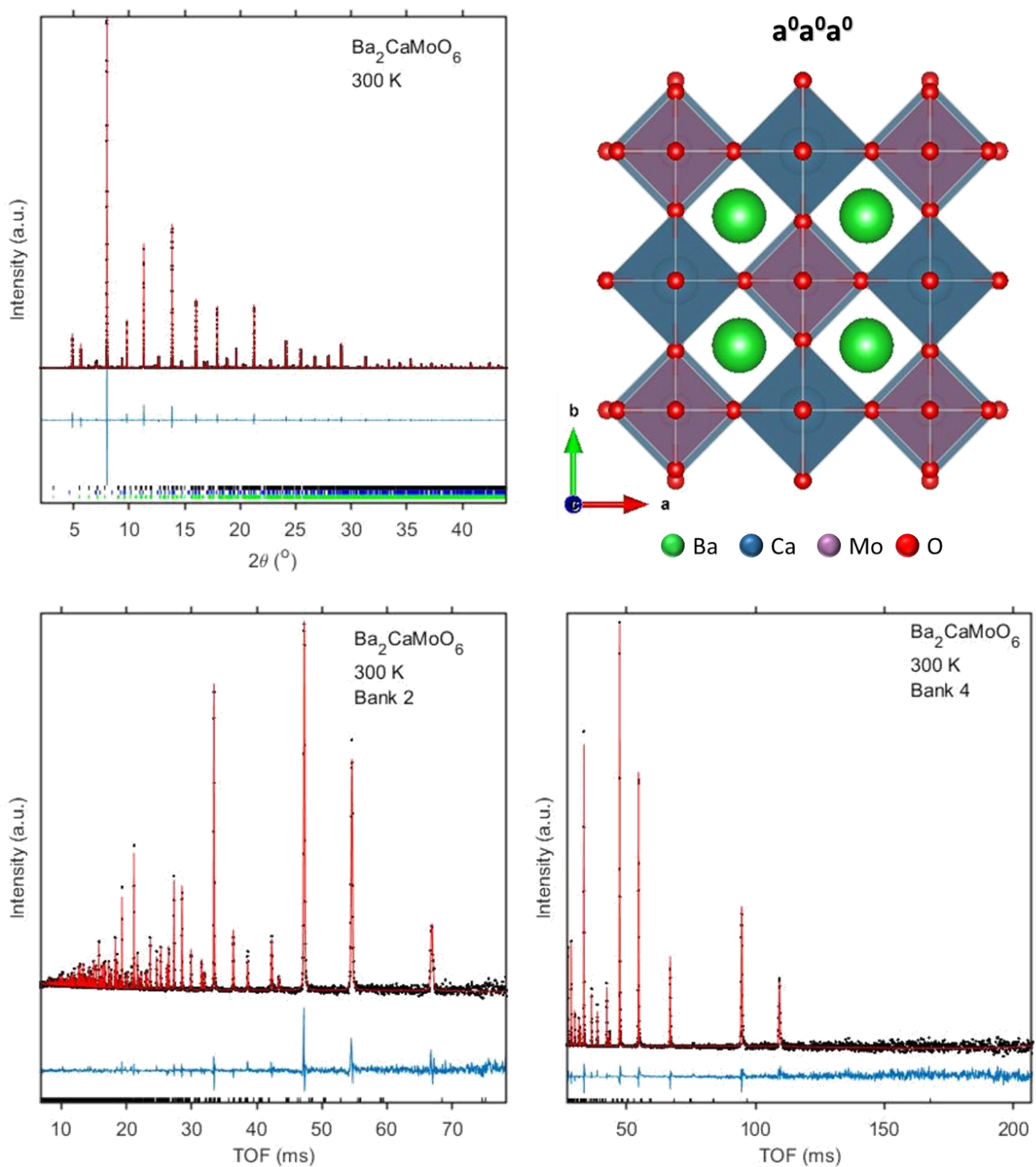

Figure 4. Rietveld refinements of 300 K Ba$_2$CaMoO$_6$ synchrotron data (top left) and neutron diffraction data (second row). Black dots are observed data, red line is calculated data, and blue line is the difference curve. Black tick marks denote the main phase, blue tick marks denote BaMoO$_4$, and green tick marks denote MoO$_3$. The cubic crystal structure is shown at the top right.

**Error! Reference source not found.** shows the Rietveld refinements of 300 K data and Figure 5 shows the Rietveld refinements of 100 K and 10 K data. Refined crystallographic and atomic parameters are in Table 1 and Table 2 respectively.

The synchrotron room temperature data does not contain additional low intensity super structure peaks, which are indicative of NCOT structures, relative to the pattern expected for the cubic structure.[11] Combined Rietveld refinements of the room temperature data is consistent with the refinement results of the laboratory XRD, which showed the prototypical cubic perovskite structure (**Error! Reference source not found.**) with a $\chi^2$ of 1.49.

In the related compound, $Ba_2CaWO_6$, a second-order phase transition from the high temperature was also observed ~220 K and was proposed to be a transition from the high temperature cubic structure, $Fm\text{-}3m$, to a tetragonal structure, $I4/m$,[19,20] or from the $I4/m$ structure to the monoclinic structure, $I2/m$.[21] Therefore, fitting of the data with both the tetragonal and the monoclinic model was attempted.

Low temperature synchrotron and neutron data were refined independently due to differences in temperatures. Rietveld refinement of the 100 K synchrotron data gave a $\chi^2$ of 3.38 and the 10 K neutron gave a $\chi^2$ of 2.12 when refining to the $I4/m$ structure. It is noted that there is a small negative displacement parameter for molybdenum in the synchrotron data, however it is considered small enough that it is not significant and could be set to zero without changing the overall refinement. A decrease in the unit cell volume of 0.3067 Å$^3$ is seen between the 100 K and the 10 K data as expected upon cooling. The $I4/m$ structure (Figure 5) shows an out of phase rotation around the $c$-axis with a Glazer notation of

$a^0a^0c^-$.[22,23] Refinements, using the monoclinic model were worse than the tetragonal model and thus it was determined that the tetragonal structure is a more accurate model to this data.

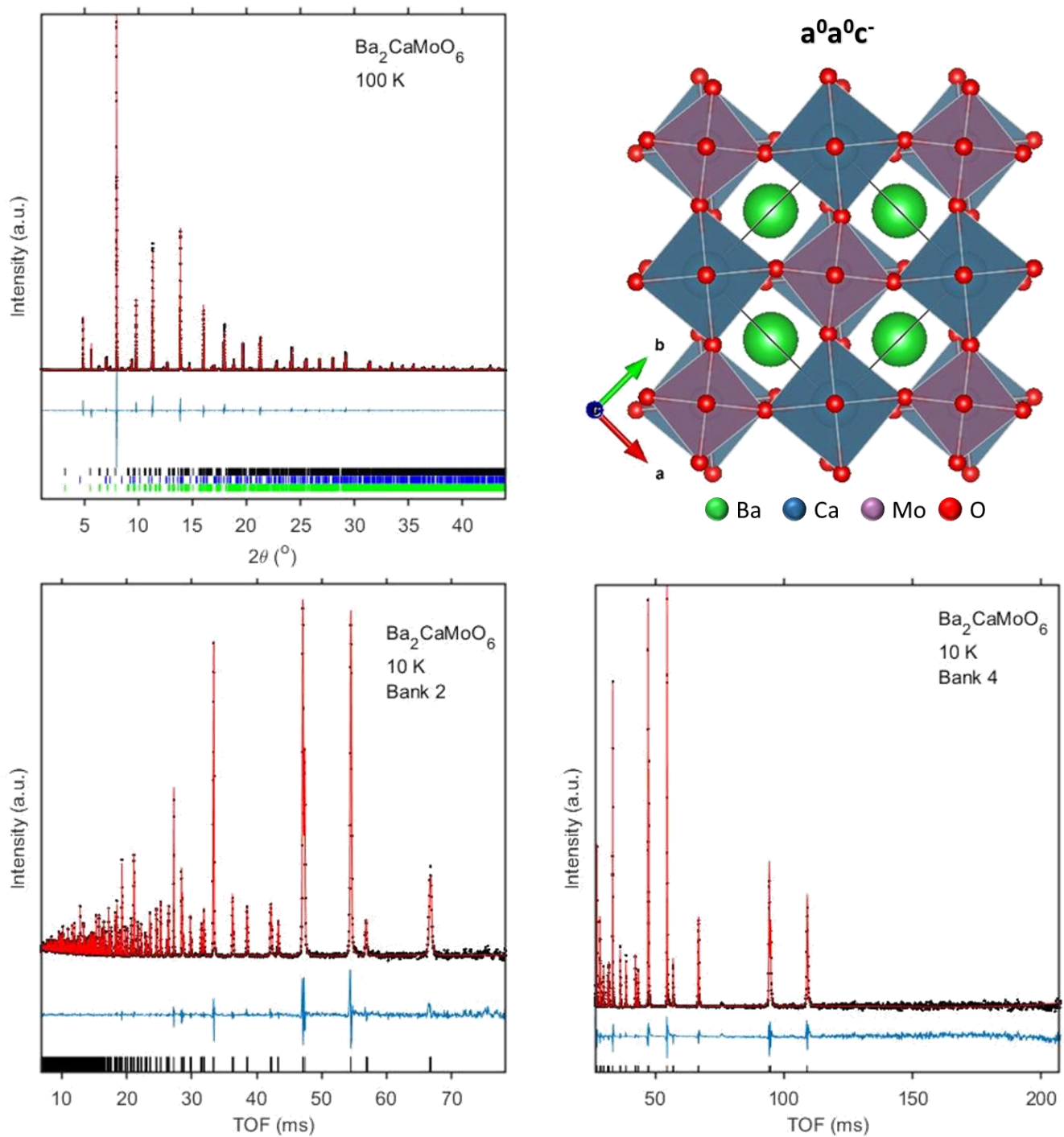

Figure 5. Rietveld refinements of 100 K Ba$_2$CaMoO$_6$ synchrotron data (top left) and 10 K neutron diffraction data (second row). Black dots are observed data, red line is calculated data, and blue line is the difference curve. Black tick marks denote the main phase, blue tick marks denote BaMoO$_4$, and green tick marks denote MoO$_3$. The tetragonal crystal structure is shown at the top right.

Table 1. Crystallographic results from TOF neutron data and synchrotron data for Ba$_2$CaMoO$_6$ at 10 K, 100 K and 300 K[b]

| Space group | I4/m (No. 87) | I4/m (No. 87) | Fm-3m (No. 225) |
|---|---|---|---|
| a (Å) | 5.90099(5) | 5.905763(6) | 8.378231(5) |
| b (Å) | 5.9010(7) | 5.905763(6) | 8.378231(5) |
| c (Å) | 8.3929(1) | 8.38817(1) | 8.378231(5) |
| V (Å$^3$) | 292.2563(5) | 292.56302(5) | 588.108(1) |
| Synchrotron λ (Å) | - | 0.412728 | 0.412728 |
| Temperature (K) | 10 | 100 | 300 |
| data type | TOF neutron | synchrotron | synchrotron and TOF neutron |
| number of reflections | 34 | 63 | 108 |
| R$_{wp}$ | 3.21 | 15.73 | 3.61 |
| $\chi^2$ | 2.12 | 3.38 | 1.49 |

Table 2. Atomic Position for Ba$_2$CaMoO$_6$ obtained from TOF neutron and synchrotron data at 10 K, 100 K and 300 K

| Atom | Wyckoff position | x | y | z | Occupancy | B |
|---|---|---|---|---|---|---|
| Ba$_2$CaMoO$_6$ TOF neutron 10 K | | | | | | |
| Ba | 4d | 0 | 0.5 | 0.25 | 1 | 0.08(2) |
| Ca | 2a | 0 | 0 | 0 | 1 | 0.17(3) |
| Mo | 2b | 0.5 | 0.5 | 0 | 1 | 0.08(2) |
| O(1) | 4e | 0 | 0 | 0.2716(3) | 1 | 0.35(1) |
| O(2) | 8h | 0.2457(3) | 0.2972(2) | 0 | 1 | 0.35(1) |
| Ba$_2$CaMoO$_6$ synchrotron 100 K | | | | | | |
| Ba | 4d | 0 | 0.5 | 0.25 | 1 | 0.288(4) |
| Ca | 2a | 0 | 0 | 0 | 1 | 0.20(2) |
| Mo | 2b | 0.5 | 0.5 | 0 | 1 | -0.039(7) |
| O(1) | 4e | 0 | 0 | 0.2711(4) | 1 | 0.23(4) |
| O(2) | 8h | 0.2464(6) | 0.2939(6) | 0 | 1 | 0.23(4) |
| Ba$_2$CaMoO$_6$ synchrotron and TOF neutron 300 K | | | | | | |
| Ba | 8c | 0.25 | 0.25 | 0.25 | 1 | 0.719(5) |
| Ca | 4b | 0.5 | 0.5 | 0.5 | 1 | 0.28(2) |
| Mo | 4a | 0 | 0 | 0 | 1 | 0.136(6) |
| O | 24e | 0.2270(2) | 0 | 0 | 1 | 0.89(3) |

[b] The goodness of fit parameters are define as follows: $\chi^2 = \frac{M}{N_{obs} - N_{var}}$ and $R_{wp} = \sqrt{\Sigma w(I_o - I_c)^2}$

To ensure the quality of the refinements the bond lengths, bond valence sums, and distortion indices in each of the structures were analyzed and are shown in Table 3. It can be seen that the cation-oxygen bond lengths are consistent with one another in the low temperature structure in which each cation adopts two different sites. Furthermore, the cation-oxygen bond lengths are consistent with expected bond lengths for the given ionic radii.[24] Analysis of the bond valence sums, $v_{ij}$, which is given by

$$v_{ij} = \exp\left[R_0 - \frac{d_{ij}}{0.37}\right]$$

where $R_0$ is the bond valence parameter and $d_{ij}$ is the bond length between atoms $i$ and $j$, shows a small deviation from the expected oxidation state for $Ba^{2+}$ and $Mo^{6+}$, but $Ca^{2+}$ consistently shows approximately 0.6 higher than expected value for all refined temperatures. The distortion index, D, is a measure of the refined deviation of the cuboctahedra and octahedra of the refined $A$ and $B$-sites respectively and is given by,

$$D = \frac{1}{n \sum_n \frac{|d_{ij} - d_{avg}|}{d_{avg}}}$$

where n is the number of bonds defining the polyhedra and $d_{avg}$ is the average bond length in the polyhedra. Given the similarities in the refined bond lengths it can be seen that the distortion indices are low for all cation sites in the low temperature structures. Given the symmetry of $A$ and $B$-sites in the cubic structure, the polyhedra cannot deviate from ideal and thus have a distortion index of zero. Overall, bond lengths, bond valence sums, and distortion indices indicate that the propose room and low temperature structures of $Ba_2CaMoO_6$ are reasonable.

Table 3. Bond valence sums and distortion index for TOF neutron and synchrotron data for $Ba_2CaMoO_6$ at 10 K, 100 K and 300 K

| Atom/Bond | TOF neutron 10 K | Synchrotron 100 K | Synchrotron and TOF neutron 300 K |
|---|---|---|---|
| Bond Length | | | |
| Ba-O1 | 2.95601(2) | 2.9582(3) | 2.96767(1) |
| Ba-O2 | 3.1193(1) | 3.107(3) | |
| Ca-O1 | 2.279(3) | 2.274(4) | 2.2755(2) |
| Ca-O2 | 2.2755(2) | 2.265(4) | |
| Mo-O1 | 1.918(3) | 1.920(4) | 1.9136(2) |
| Mo-O2 | 1.9194(2) | 1.930(4) | |
| Bond Valence Sum | | | |
| Ba | 2.021 | 2.005 | 1.896 |
| Ca | 2.599 | 2.660 | 2.606 |
| Mo | 5.811 | 5.690 | 5.894 |
| O | 2.351 | 1.749 | 2.049 |
| Distortion Index | | | |
| Ba | 0.03489 | 0.03208 | 0 |
| Ca | 0.00062 | 0.00177 | 0 |
| Mo | 0.00037 | 0.00228 | 0 |

## Conclusions

In this study we found that $Ba_2CaMoO_6$ does not adopt a NCOT structure between 10 K and 300 K but instead adopts the same cooperative octahedral tilted structure, *I*4/*m*, below the transition temperature of 200 K as also seen in the tungsten analog. Above the transition temperature, it adopts the cubic perovskite structure. This indicates that having a difference of *B* and *B'* in the expected range of 0.50 and 0.66 Å and/or a small ΔIR of counter ions may play a significant role in the formation of NCOT compounds. Additional, studies to isolate these parameters from one another are underway. It should be noted that to date no known NCOT oxide with $Mo^{6+}$ on the *B'* has been observed and thus the role of the covalency of the *B'*-site in the formation of NCOT compounds is also still under investigation.

## Acknowledgments

The authors would like to acknowledge start-up funds from California State University, Fullerton. They would like to thank Dr. Tyrel McQueen, Dr. Zachary Kelly, and Dr. Benjamin Trump for collecting low temperature powder x-ray diffraction data on the Bruker D8 supported by the Sloan Foundation. This research used resources of the Advanced Photon Source, a U.S. Department of Energy (DOE) Office of Science User Facility operated for the DOE Office of Science by Argonne National Laboratory under Contract No. DE-AC02-06CH11357. This research also used resources at the Spallation Neutron Source, a DOE Office of Science User Facility operated by the Oak Ridge National Laboratory.

**References**

[1]   J. Iñiguez, First-principles approach to lattice-mediated magnetoelectric effects, Phys. Rev. Lett. 101 (2008) 277–285. doi:10.1103/PhysRevLett.101.117201.

[2]   G. Shirane, Neutron scattering studies of strucutral phase transitions at Brookhaven, Rev. Mod. Phys. 46 (1974) 437–449.

[3]   G.H. Kwei, S.J.L. Billinge, S.W. Cheong, J.G. Saxton, Pair-distribution functions of ferroelectric perovskites: Direct observation of structural ground states, Ferroelectrics. 164 (1995) 57–73. doi:10.1080/00150199508221830.

[4]   M.B. Smith, K. Page, T. Siegrist, A. Et, Crystal structure and the paraelectric-to-ferroelectric phase transition of nanoscale $BaTiO_3$, J. Am. Chem. Soc. 130 (2008) 6955–6963.

[5]   N. Choudhury, E.J. Walter, a. I. Kolesnikov, C.-K. Loong, Origin of the large phonon band-gap in $SrTiO_3$ and the vibrational signatures of ferroelectricity in $ATiO_3$ perovskite: First principles lattice dynamics and inelastic neutron scattering of $PbTiO_3$, $BaTiO_3$ and $SrTi$, (2008) 11. doi:10.1103/PhysRevB.77.134111.

[6]     E.I. Science, Calculation and Analysis of the Dielectric Functions for $BaTiO_3$ , $PbTiO_3$, and $PbZrO_3$, 51 (2013) 532–539. doi:10.6122/CJP.51.532.

[7]     T. Trautmann, C. Falter, Lattice dynamics, dielectric properties and structural instabilities of $SrTiO_3$ and $BaTiO_3$, J. Phys. Condens. Matter. 16 (2004) 5955–5977. doi:10.1088/0953-8984/16/32/028.

[8]     Q. Zhang, T. Cagin, W. a Goddard, The ferroelectric and cubic phases in $BaTiO_3$ ferroelectrics are also antiferroelectric., Proc. Natl. Acad. Sci. U. S. A. 103 (2006) 14695–14700. doi:10.1073/pnas.0606612103.

[9]     D. Khatib, R. Migoni, Lattice dynamics of $BaTiO_3$ in the cubic phase, J. Phys. …. 1 (1999) 9811. doi:10.1088/0953-8984/1/49/002.

[10]    G. King, A.M. Abakumov, J. Hadermann, A.M. Alekseeva, M.G. Rozova, T. Perkisas, P.M. Woodward, G. Van Tendeloo, E. V. Antipov, Crystal structure and phase transitions in Sr3WO6, Inorg. Chem. 49 (2010) 6058–6065. doi:10.1021/ic100598v.

[11]    A.M. Fry, P.M. Woodward, Structures of α-$K_3MoO_3F_3$ and α-$Rb_3MoO_3$ $F_3$: Ferroelectricity from Anion Ordering and Noncooperative Octahedral Tilting, Cryst. Growth Des. 13 (2013) 5404–5410. doi:10.1021/cg401342q.

[12]    X. Tan, P.W. Stephens, M. Hendrickx, J. Hadermann, C.U. Segre, M. Croft, C.J. Kang, Z. Deng, S.H. Lapidus, S.W. Kim, C. Jin, G. Kotliar, M. Greenblatt, Tetragonal $Cs_{1.17}In_{0.81}Cl_3$: A Charge-Ordered Indium Halide Perovskite Derivative, Chem. Mater. (2019). doi:10.1021/acs.chemmater.8b04771.

[13]    F. Javier Zúñiga, A. Tressaud, J. Darriet, The low-temperature form of $Rb_2KCrF_6$ and $Rb_2KGaF_6$: The first example of an elpasolite-derived structure with pentagonal bipyramid in the B-sublattice,

J. Solid State Chem. 179 (2006) 3607–3614. doi:10.1016/j.jssc.2006.07.032.

[14] A.M. Abakumov, M.D. Rossell, A.M. Alekseeva, S.Y. Vassiliev, S.N. Mudrezova, G. Van Tendeloo, E. V. Antipov, Phase transitions in $K_3AlF_6$, J. Solid State Chem. 179 (2006) 421–428. doi:10.1016/j.jssc.2005.10.044.

[15] A.M. Abakumov, G. King, V.K. Laurinavichute, M.G. Rozova, P.M. Woodward, E. V. Antipov, The Crystal Structure of α-$K_3AlF_6$: Elpasolites and Double Perovskites with Broken Corner-Sharing Connectivity of the Octahedral Framework, Inorg. Chem. 48 (2009) 9336–9344. doi:10.1021/ic9013043.

[16] M.T. Anderson, K.B. Greenwood, G.A. Taylor, K.R. Poeppelmeier, B-Cation arrangements in double perovskites, Prog. Mater. Sci. 22 (1993) 197–233.

[17] B. AXS, Topas Academic General Profile and Structure Analysis Software for Powder Diffraction Data, Karlsruhe, Ger. (2004).

[18] J. Terzic, J.C. Wang, F. Ye, W.H. Song, S.J. Yuan, S. Aswartham, L.E. Delong, S. V. Streltsov, D.I. Khomskii, G. Cao, Coexisting charge and magnetic orders in the dimer-chain iridate $Ba_5AlIr_2O_{11}$, Phys. Rev. B - Condens. Matter Mater. Phys. 91 (2015) 6–11. doi:10.1103/PhysRevB.91.235147.

[19] W.T. Fu, Y.S. Au, S. Akerboom, D.J.W. IJdo, Crystal structures and chemistry of double perovskites $Ba_2M(II)M′(VI)O_6$ (M=Ca, Sr, M′=Te, W, U), J. Solid State Chem. 181 (2008) 2523–2529. doi:10.1016/j.jssc.2008.06.024.

[20] W.T. Fu, S. Akerboom, D.J.W. IJdo, Crystal structures of the double perovskites $Ba_2Sr_{1-x}Ca_xWO_6$, J. Solid State Chem. 180 (2007) 1547–1552. doi:10.1016/j.jssc.2007.03.008.

[21] K. Yamamura, M. Wakeshima, Y. Hinatsu, Structural phase transition and magnetic properties of double perovskites Ba2CaMO6 (M=W, Re, Os), J. Solid State Chem. 179 (2006) 605–612. doi:10.1016/j.jssc.2005.10.003.

[22] P.M. Woodward, Octahedral Tilting in Perovskites. I. Geometrical considerations, Acta Crystallogr. B. 53 (1997) 44–66. doi:10.1107/S0108768196012050.

[23] P.M. Woodward, Octahedral Tilting in Perovskites. II. Structure Stabilizing Forces, Acta Crystallogr. B. 53 (1997) 44–66. doi:10.1107/S0108768196012050.

[24] R.D. Shannon, Revised effective ionic radii and systematic studies of interatomic distances in halides and chalcogenides, Acta Crystallogr. Sect. A. 32 (1976) 751–767. doi:10.1107/S0567739476001551.